# Multicaloric effect in multiferroic $Y_2CoMnO_6$

J. Krishna Murthy and A. Venimadhav[a)]

[1]Cryogenic Engineering Centre, Indian Institute of Technology, Kharagpur-721302, India

## Abstract

We have investigated multiple caloric effects in multiferroic $Y_2CoMnO_6$. Polycrystalline sample prepared by solid state method has shown a ferromagnetic Curie temperature ~ 75 K with second order phase transition; a maximum magneto entropy change ($-\Delta S_M^{max}$) of ~ 7.3 J $kg^{-1}$ $K^{-1}$ with reasonable relative cooling power ~ 220 J/kg is found without thermal and magnetic hysteresis loss. Electric field driven entropy change ($-\Delta S_E^{max}$) of ~ 0.26 J $m^{-3}$ $K^{-1}$ obtained using the Maxwell's relation and estimated magnetically induced total temperature change of ~ 5.45 K around Curie temperature that confirms the multicaloric effect in the sample.

## 1. Introduction

Solid state cooling phenomena based on adiabatic relaxation of an order parameter (magnetic, electric and elastic) to produce a temperature reduction of the solid refrigerant is known for decades. Magnetocaloric technology was developed for milli Kelvin temperatures as an alternative to conventional vapor compression refrigeration [1-3]. A wide variety of magnetic materials are accessible for magnetocaloric applications depending on the desired temperature range of operation. For example, Gd and its alloys [4, 5], Mn based intermetallic compounds [6], and manganite systems [7] are good for room temperature refrigeration applications, while for low temperature refrigeration purpose ($LH_2$ – liquid hydrogen and ultra-low temperature (mK-μK)), molecular magnetic systems [8] and paramagnetic (PM) rare earth salts [2, 9, 10] are suitable. The electrocaloric effect (ECE) is still in its infancy and best results are reported in either inorganic $BaTiO_3$ or derivatives of polymer ferroelectric (PVDF) materials are suitable only for close to room temperature applications [11, 12]. By analogy, barocaloric effect was found in structural phase transformation systems, like shape memory alloys [13]; a large amount of latent heat can be released by external pressure.

Recent discoveries of giant caloric effect in ferroic materials have opened door to commercial applications at room and cryogenic temperatures; additionally, this method offers clean and energy efficient technology [1, 11, 12]. The prospects of solid state technology will be high with novel multiple caloric materials where one or more order parameters can produce cooling to achieve larger temperature gradient [14-18]. Multiferroics, facilitate multifunctional properties like ferromagnetic (FM), ferroelectricity and ferroelasticity with a possibility to control various order parameter(s) with respective or complementary external fields [19, 20]. In multiferroic systems, since there is a large change in the magnetization and polarization near to their ferroic ordering temperatures, they can be expected to show multicaloric effects in terms of magnetocaloric effect (MCE) and ECE. The key parameters to assess the solid state refrigerant materials are: (i) high $\Delta S_M$, RCP and $\Delta T_{ad}$, (ii) absence of thermal and magnetic hysteresis near to the transition temperature and (iii) good thermal conductivity to transfer heat from the system (to be cooled) to the refrigerant.

MCE has been investigated in a number of multiferroic systems. Improper magnetic multiferroics are of interest due to the presence of both ferromagnetic and ferroelectric properties with magneto-electric coupling [21]. Recently, multiferroic behaviour has been reported in $Y_2CoMnO_6$ (YCMO) double perovskite system with simultaneous presence of magnetic and electric order [22]. In this report, we have studied temperature dependence of MCE and ECE properties in YCMO system under magnetic and electric fields respectively; these values were compared with other magnetic multiferroic systems and are shown in the Table. 1.

## 2. Experimental details

Polycrystalline YCMO ceramic sample was prepared by conventional solid state reaction method; crystal structure and lattice parameters were consistent with previous reports [22]. Temperature and magnetic field dependence of dc and ac susceptibility measurements were done using EverCool Quantum Design SQUID-VSM magnetometer. The temperature dependent specific heat data was carried out using Quantum design-physical property measurement system. Temperature dependent pyroelectric current (I) was measured by using Keithley-6514A electrometer.

## 3. Results and discussion

*3.1. Magnetization and specific heat study*

Fig. 1(a) shows the temperature (T) variation of magnetization (M) in zero-field cooled (ZFC) and field-cooled warming (FCW) modes with 0.01 T dc field; as decreasing the temperature from 300 K down to 2 K, a PM to FM ordering is observed around 75 K, followed by this a broad shoulder around 55 K in ZFC magnetization can be noticed. The Curie-Weiss (CW) law fit to the FCW magnetization data (inset to Fig. 1(a)) has shown CW temperature (θ) ~ 74.7 K and the effective PM moment of ~ 7.24 $\mu_B$/f.u., which is close to the theoretically calculated effective magnetic moment of Co and Mn ions, $\mu_{eff}^2 = \mu_B^2[(g^2S(S+1))_{Co}+(g^2S(S+1))_{Mn}] = 7.14\mu_B$/f.u.. Here, the spin (S) only value of both $Co^{2+}$ and $Mn^{4+}$ magnetic ions is 3/2 and the Lande's splitting factor (g) for Mn is 2 while for Co ions it is 3.1 due to its orbital contribution [31]. Further, isothermal magnetic field (H) dependent magnetization measurements (i.e., M (H)) at 5 K and 75 K are shown in the Fig.1 (b).There is no noticeable hysteresis at the transition temperature; observation of large remanent magnetization and coercive field ($H_C$) ~ 1.12 T at low temperature are the typical features of rare earth Co-Mn double perovskite systems [32, 33].

Temperature variation of specific heat ($C_P$) (Fig. 2(a)) has shown a λ-like peak (characteristic of second order phase transition [5]) at 75 K and this peak is attributed to the absorption of heat exploited in randomization of spins around $T_C$. However, no anomaly is noticed around 55 K. The low temperature (T < 30 K) $C_P$ data can be fitted (as shown in the lower inset to the Fig. 2(a)) to the expression,

$$C_P (T) = \gamma T + \beta T^3 \quad (1)$$

Here the first and the second term on right side describe the electronic and phonon contributions to $C_P$ (T) respectively. The linear coefficient γ can be attributed to the charge carriers and is proportional to the density of the states at the fermi level and the coefficient β is related to Debye temperature ($\Theta_D$). From the fitting, $\gamma = 16.85 \pm 3$ m J/mol K$^2$ and $\beta = 0.56 \pm 0.014$ m J/mol K$^4$ are found and the small value of $\gamma$ suggests the highly insulating nature of YCMO. From the lattice part of $C_P$, the Debye temperature of ~ 324 K has been calculated from $\Theta_D = (\frac{12\pi^3 PR}{5\beta})^{\frac{1}{3}}$ value, where, $R$ is universal gas constant and $P$ is the number of atoms in each unit cell.

The frequency independent in - phase component ($\chi'(T)$) of ac susceptibility (ACS) at 75 K confirms the FM ordering as shown in Fig. 2(b). While the frequency dispersion around 55 K suggests the slow dynamics of domain wall motion or spin glass nature similar to the Lu$_2$CoMnO$_6$ multiferroic system [32]. To verifying such an ambiguity, we have performed the memory effect in ACS at low frequencies (~ 300 m Hz) with an ac field of ~ 0.5 Oe using the standard protocol reported elsewhere [34]. The susceptibility data is recorded during the heating mode with a halt time (~ 3 hours at 50 K) denoted as $\chi'_{halt}$ and without halt time denoted as $\chi'_{ref}$. The difference between these two curves, i.e., ($\Delta\chi' = \chi'_{halt} - \chi'_{ref}$) with temperature is shown in Fig. 2(c). Here, absence of dip in $\Delta\chi'$ vs. T at the halt temperature (~ 50 K) indicates the absence of spin glass behaviour. Further, we have measured the out-of phase ($\chi''$) component of ACS with low ac driving field of ~ 0.5 Oe and 623 Hz to verifying the thermal hysteresis around the magnetic anomaly (~ 55 K). During this measurement, the data is taken in temperature stabilization mode with sufficient time to equilibrate spins. As shown in Fig. 2(d), the heating and cooling data of ACS curves exactly coincide at T$_C$ and indicates the transition is of a second order kind, while a significant thermal hysteresis is observed around 55 K. We have performed this measurement on both powder and sintered pellet forms of YCMO and obtained the same behaviour. This rules out the possibility of pinning of domain wall motion at the grain boundaries and it is indeed intrinsic nature of the YCMO [35].

*3.2 Magnetocaloric effect*

The MCE has been estimated from the isothermal initial magnetization data using Maxwell relation,

$$\Delta S_M(T,H) = \int_{H_i}^{H_f} \left(\frac{\partial M}{\partial T}\right)_H dH \qquad (2)$$

In YCMO, we have observed a significant thermal hysteresis and also magnetic hysteresis around 55 K below $T_C$. Because of metastablity the measured MCE using isothermal magnetization method (Eqn. (2)) will be erroneous due to the remanent magnetization for $T < T_C$. This can be avoided by measuring the isothermal M (H) curves after cooling the sample from the PM state to measured temperature (for $T < T_C$) [36]. Fig. 3(a) shows a series of isothermal initial M (H) curves in the vicinity of $T_C$ with closed temperature intervals (~ 2 K) after removing the demagnetization effects. With the Arrott plots (not shown here) we have obtained critical exponents $\delta = 3.12$, $\beta = 0.49$ and $\gamma = 1.03$ which are consistent with the mean field theory and the exchange interaction is of long-range type. In fact from the temperature dependent neutron scattering measurements, the magnetic exchange interaction in $YCo_{0.5}Mn_{0.5}O_3$ system was explained satisfactorily by mean field model [37].

Fig. 3(b) shows the temperature dependent MCE for different fields as obtained from the initial M (H) curves using the numerical expression of Eqn. (2). Maximum value of $\Delta S_M$ (H, T) at 75 K for 7 T is ~ 7.4 J/kg K and it is smaller than the theoretically calculated spin value of $|\Delta S_M| = N K_B \ln(2S+1) \sim 33$) by 4.5 times; where $N$ is the total number of spins, S is the total spin quantum number. An incomplete saturation and low value of magnetization (~ 4

$\mu_B$/f.u. even at 7 T) near to 75 K can account for the low $|\Delta S_M|$. Since there is no magnetic hysteresis at ~75 K, we could observe identical values of $|\Delta S_M|$ in both field increasing and decreasing branches of M (H) curves. The relative cooling power (RCP = $\Delta S_M^{peak}$ x $\delta T_{FWHM}$, FWHM is full width at half maxima of $|\Delta S_M|$ peak), an important parameter which indicates about the efficiency of magnetic refrigerant to transfer the heat between cold and heat reservoirs. In Fig. 3(c), we have presented the variation of RCP and $|\Delta S_M|$ of the material with magnetic field. The maximum value of RCP for a field change of 7 T is estimated to be 225 J/kg. The obtained values of $\Delta S_M^{max}$ and RCP of the present compound are comparable to those observed in manganites [38]. Therefore, the magnetocaloric nature of YCMO satisfies the important criteria of magnetic refrigerant such as moderate value of magnetic entropy change, high cooling efficiency and absence of both thermal and magnetic hysteresis.

*3.3 Electrocaloric effect*

YCMO being a magnetic multiferroic material is expected to show ECE. Since improper multiferroics show small electric polarization, the conventional polarization vs. electric field (P-E) hysteresis loop method may not be suitable to characterizing the ferroelectricity. Alternatively, the pyroelectric current method is useful to evaluate ferroelectricity [39]. The ECE can be calculated by measuring the pyroelectric current data under different polling fields. Initially, the systems is cooled under different poling fields and then sample contacts has been shorted for two hours to remove the extrinsic effects induced by the dipoles reorientation (or) release of charges from the localized states. Then pyroelctric current is measured during the heating mode with a constant temperature ramping rate of 3 K/min. Fig. 4 shows the temperature dependence of pyroelectric current and polarization data under applied poling fields. Here, the temperature dependent polarization data is obtained from pyroelectric current data with respect to time and it follows the magnetization below $T_C$. Further using Maxwell relation; $(\frac{\partial S}{\partial E})_T = (\frac{\partial P}{\partial T})_E$ [40], the ECE can be estimated as follows,

$$|\Delta S_E| = \sum [(\frac{\partial P}{\partial T})_{E_i} + (\frac{\partial P}{\partial T})_{E_f}] \times \frac{1}{2}(\Delta E) \qquad (3)$$

Here, $E_i$ and $E_{i+1}$ are the minimum and maximum poling electric fields. In the present case it is found that for the electric field at 450 kV/m the system breaks down, therefore a maximum poling field of ($E_{i+1}$) ~ 400 kV/m has been set. Inset (a) of Fig.(4) shows the temperature dependence of $|\Delta S_E|$ and it appears below the magnetic ordering with a maximum value of ~ 0.26 J m$^{-3}$K$^{-1}$ at $T_C$. Further, the temperature dependent electric polarization at 0 T and 5 T magnetic fields at a fixed poling field E ~ 200 kV/m are shown in the inset of Fig 4(b). The polarization obtained at 10 K for 0 T is ~ 82 μ C/m$^2$ suppressed to ~ 73 μ C/m$^2$ with 5 T and such a small effect with magnetic field on polarization can be ascribed to the non-saturation found in the magnetization. We have calculated the magneto-electric (ME) susceptibility (i.e., $\alpha_{ME}$ = (P (H) - P (0))/ΔH at 5 K and 75 K and values are ~ -1.8 μ C/m$^2$ T and ~ -0.327 μ C/m$^2$ T respectively. It suggests that the ME coefficient in YCMO is low. Nevertheless, the present method is promising to investigate the multicaloric effect in high ME materials $Co_4Nb_2O_9$ and $CaMn_4O_7$ [41, 42].

In the present study we have estimated the MCE and ECE individually in this YCMO multiferroic system. Further, an additional caloric coupling term due to multuiferroicity can be estimated as [15],

$$\Delta T_m = -\frac{T}{C} \int_{H_i}^{H_f} [(\frac{\partial M}{\partial T})_{H,E} + \frac{\alpha_m}{\varepsilon_0 \chi^e}(\frac{\partial P}{\partial T})_{H,E}] d H \qquad (4)$$

Here $C_P$ is the specific heat (6.230 x $10^5$ J $m^{-3}$ $K^{-1}$) at T ~75 K, $\chi^e$ is the electrical susceptibility ($\varepsilon_r - 1$ = 16.63), $\alpha_m$ is the ME coupling parameter (-0.327 x$10^{-6}$ C $m^{-2}$ T) and H is the applied magnetic field which is 5 T. From the temperature and magnetic field dependent magnetization and polarization we have obtained $\frac{\partial M}{\partial T}$ = 1.5908 x $10^{-3}$ emu $mol^{-1}$ $K^{-1}$ and $\frac{\partial P}{\partial T}$ = 0.32 x $10^{-6}$ C $m^{-2}$ $K^{-1}$ respectively. And total adiabatic temperature change due to magnetically induced multicaloric effect is ($\Delta T_m$) ~ 5.45 K per each cycle. Here, the standard MCE contribution is ~ 5.79 K and the contribution due to multicaloric coupling term is ~ 0.34 K to the total temperature change. Since $\alpha_m$ has negative sign (i.e., the decreasing of electric polarization under magnetic field), the resultant total adiabatic temperature change decreases due to the ME coupling. It suggests that to take advantage of multiferroic effect for cooling application, one would need either conventional magnetocaloric effect with positive ME coupling or inverse magnetocaloric with negative ME coefficient. Magnetically induced ferroelectric systems like $Cr_2O_3$ and $Co_4Nb_2O_9$, are attractive as they show linear ME effect with positive sign near $T_C$.

## 4. Conclusions

In summary, the magnetization and specific heat study confirms the second order phase transition of FM order in YCMO. Absence of thermal hysteresis and magnetic hysteresis loss near to $T_C$, reasonable values of $\Delta S_M$ and RCP signifies YCMO as magnetic refrigerant at cryogenic temperatures. The simultaneous appearance of ECE and MCE at $T_C$ in YCMO illustrate the potential of magnetic multiferroics for multicaloric applications. Magnetic field induced multicaloric effect has been estimated and the present study signifies the sign of ME coupling on the cooling property.

**Acknowledgments:**

We would like to thank Prof. Melvin M Vopson for fruitful discussions. The authors acknowledge DST, FIST in Cryogenic Engineering Centre and IIT Kharagpur for funding VSM-SQUID magnetometer. Krishna thanks CSIR-UGC, Delhi for SRF.

**Table. 1:** Transition temperature, $|\Delta S_M^{max}|$ and RCP under magnetic field in various magnetic multiferroic systems (except $BiFeO_3$ which is proper multiferroic system), here $T_0$ is the temperature of maximum of $|\Delta S_M|$.

| Multiferroic material | $T_0$ (K) | $|\Delta S_M^{max}|$ (J/kg K) | RCP (J/kg) | $|\Delta S_E^{max}|$ (J/m³ K) | Ref. |
|---|---|---|---|---|---|
| $BiFeO_3$ | 18 K | 8.4 @ 8 T | ___ | ___ | 23 |
| $RCrO_4$ (R= Dy and Ho) | 18-23 K | 28-30 @ 7 T | 580 - 740 | ___ | 24 |
| $RMnO_3$ (R= Dy, Tb, Ho and Yb) | 5-15 K | 2 - 5.5 (J/mol K) @ 8 T | 160-25 (J/mol) | ___ | 25 |
| $DyCrO_3$ | 15 K | 8.4 @ 4 T | 217 | ___ | 26 |
| $YFe_{1-x}Mn_xO_3$ (x=0.4) | 300-370 K | 2-5 @ 7 T | ___ | ___ | 27 |
| $DyFe_{0.5}Cr_{0.5}O_3$ | 5 K | 11.3 @ 4.5 T | ___ | ___ | 28 |
| $HoMn_2O_5$ | 17 K | 7.8 @ 7 T | 216.7 | ___ | 29 |
| GdMnO$_3$ | 20 K | | | 0.73 @ 4.23kV/cm (extracted from Fig. 3(a) of Ref. 30) | 30 |
| Y$_2$CoMnO$_6$ | 75 K | 7.5 @ 7 T | 225 | 0.26 @ 2 kV/cm | This work |

**Figure captions:**

**Fig. 1** (a): Temperature (T) dependent magnetization (M) under ZFC and FCW modes with 0.01 T field, inset is the CW law fit to the experimental data of ($\chi^{-1}_{dc}$) vs. *T*, (b): isothermal M (H) measurements at 5 K and 75 K.

**Fig. 2** (a): Temperature (T) dependent heat capacity $C_P$ (T) data and inset shows low temperature $C_P$ data fitted to the Eqn. (1), (b): T dependence of $\chi'$ for different frequencies, (c): T variation of $\Delta\chi'$ (=$\chi'_{halt}$ - $\chi'_{ref}$) around the halting temperature and (d): T dependent $\chi''$ for an ac field of 0.5 Oe and 623 Hz under cooling and heating modes.

**Fig. 3**: (a): Magnetic fields (H) dependence of isothermal M (H) curves for selected temperatures in the vicinity of T$_C$ with 2 K temperature interval, (b): Temperature (T) dependence of |$\Delta S_M$| for different magnetic fields, and (c): The field dependent peak values of |$\Delta S_M$| and RCP.

**Fig. 4**: Temperature (T) variation of pyroelectric current (*I*) (left side) and polarization data (right side), inset (a) is the temperature dependent |Δ$S_E$| and (b) shows the temperature variation of polarization for 0 T and 5 T magnetic fields.

Fig 1

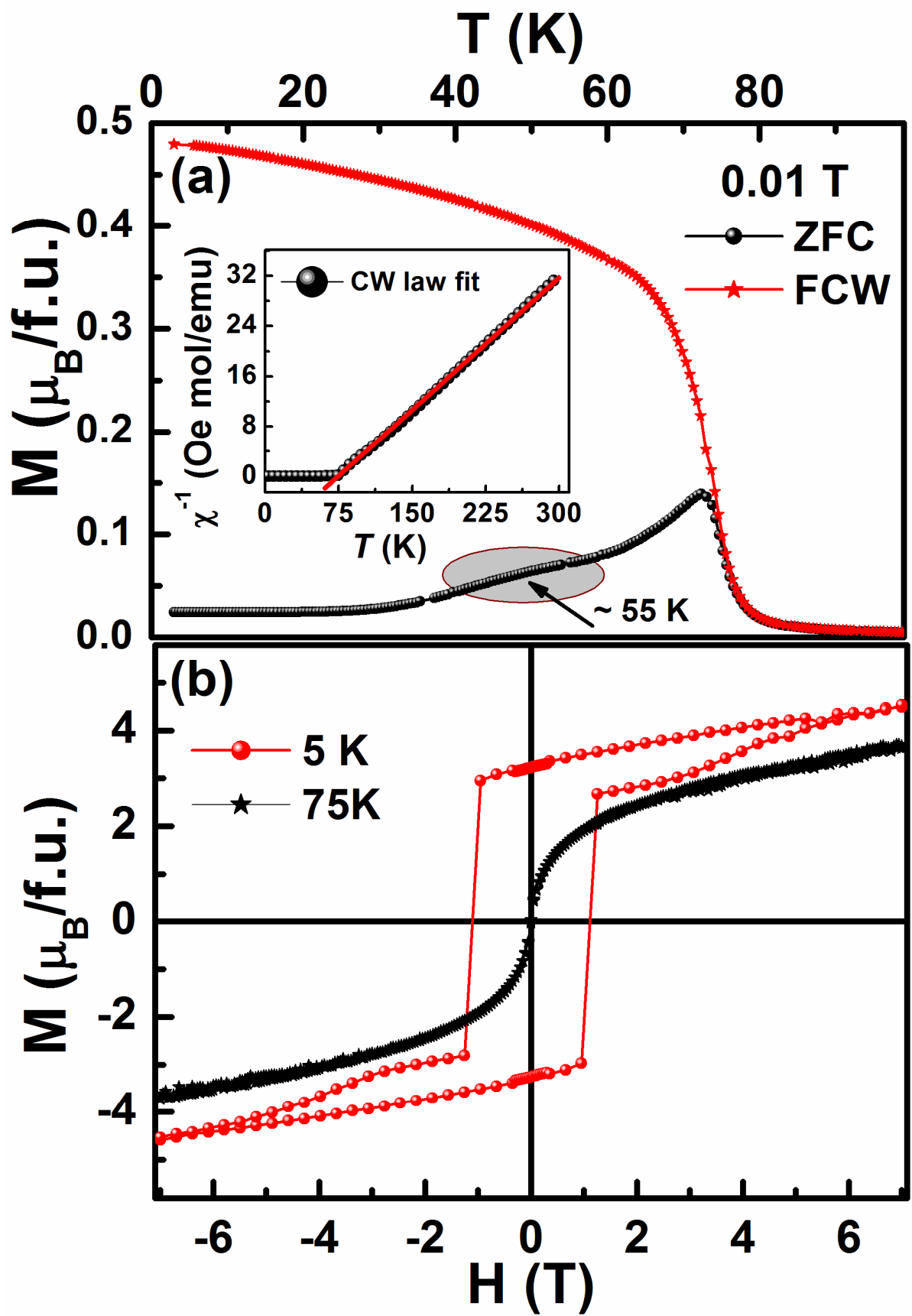

Fig 2

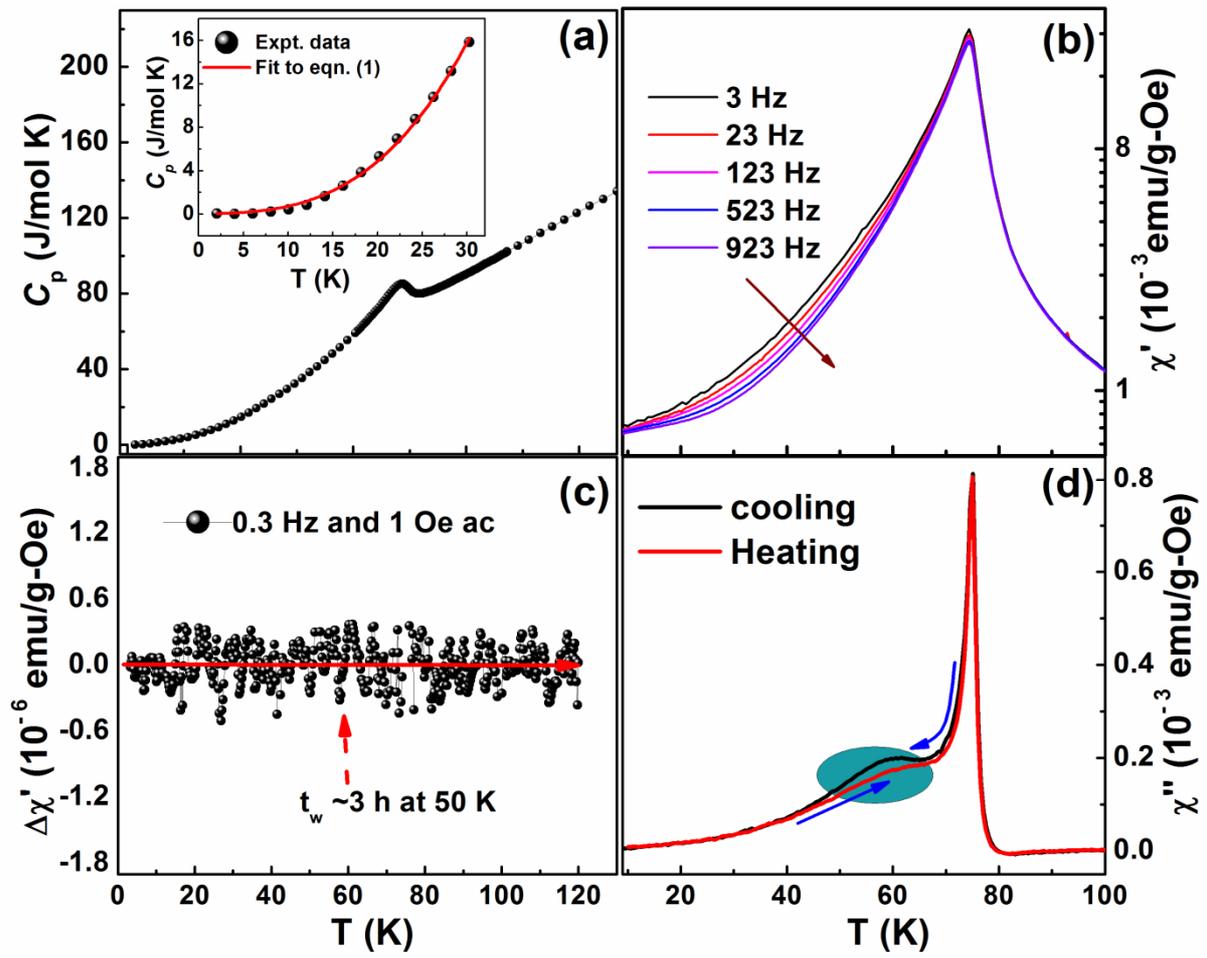

Fig 3

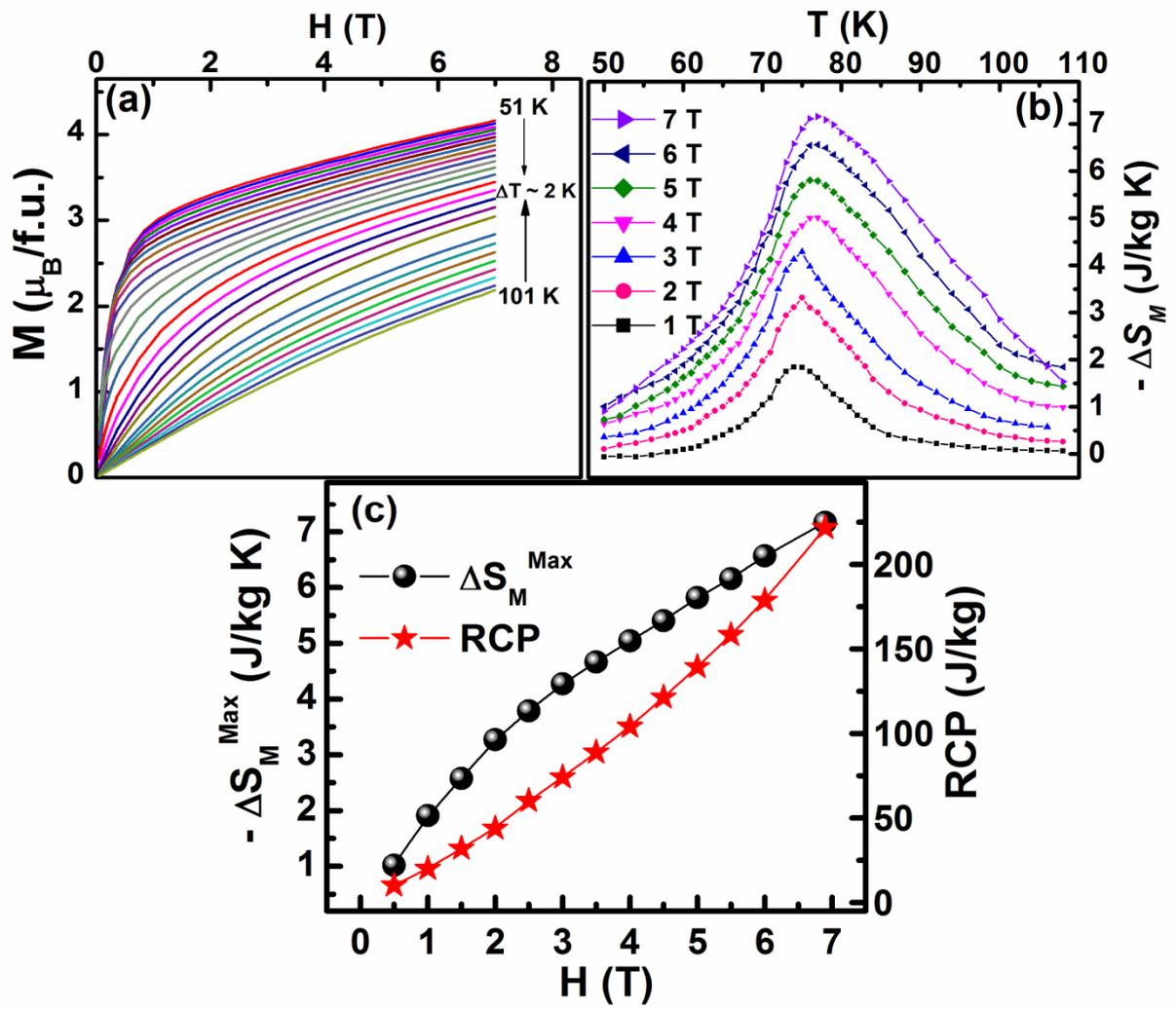

Fig. 4

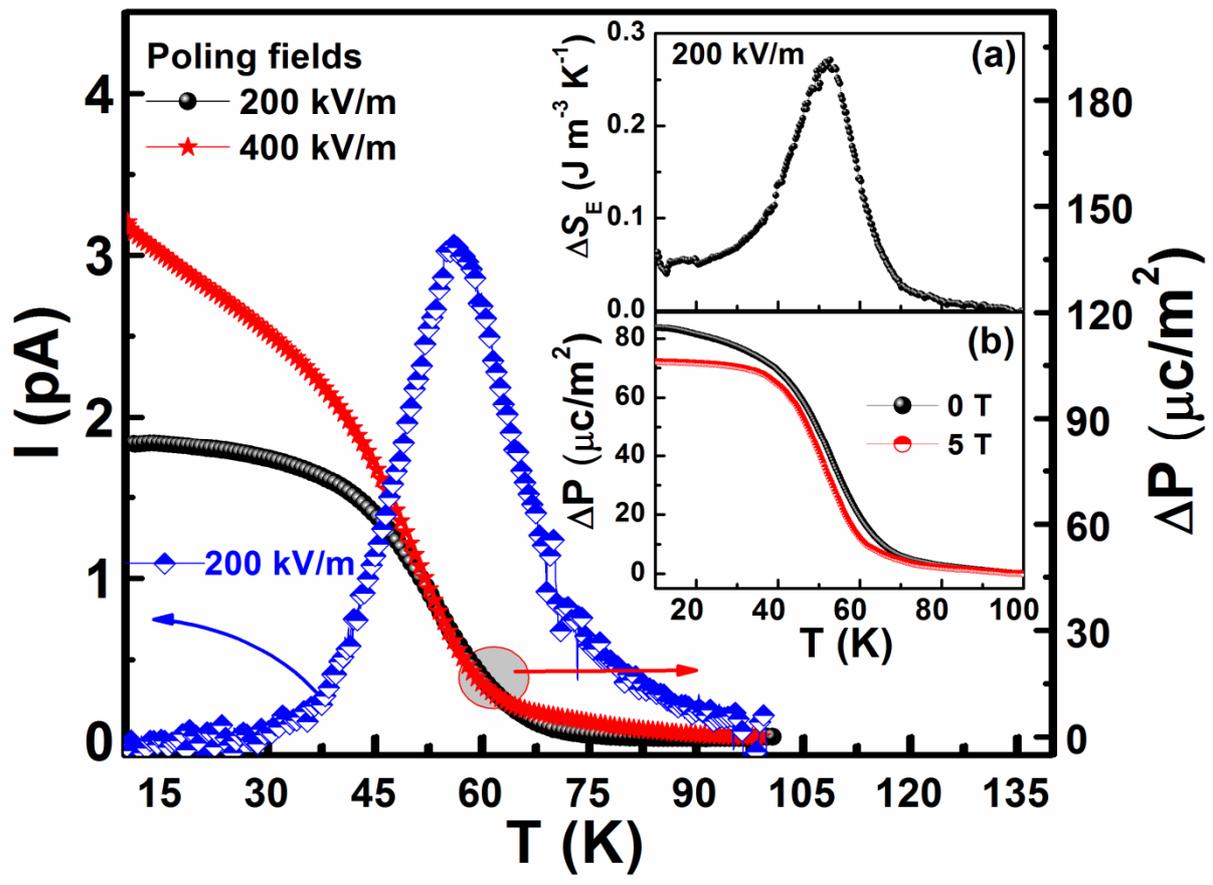